\documentclass[showpacs,pre,aps,superscriptaddress,floatfix]{revtex4}
\usepackage{amsmath}
\usepackage{amssymb,mathrsfs}
\usepackage{graphicx}
\usepackage[dvipsnames]{xcolor}

\usepackage[utf8]{inputenc}
\usepackage[english]{babel}

\newcommand{\prt}{\partial}

\newcommand{\eps}{\varepsilon}

\begin{document}
\title{Dynamics of ring solitons in an expanding cloud of a Bose-Einstein condensate}

\author{A. M. Kamchatnov}
\email{kamch@isan.troitsk.ru}
\affiliation{Institute of Spectroscopy Russian Academy of Sciences, Troitsk,
Moscow, 108840, Russia}
\affiliation{Skolkovo Institute of Science and Technology, Skolkovo, Moscow, 143026, Russia}
\affiliation{ Higher School of Economics, Physical Department, 20 Myasnitskaya ulica, Moscow 101000, Russia}
\author{B. I. Suleimanov}
\email{bisul@mail.ru}
\affiliation{Institute of Mathematics and Computing Center, Ufa Scientific
Center, Russian Academy of Sciences, ul. Chernyshevskogo 112,
Ufa, 450008 Bashkortostan, Russia}
\author{E. N. Tsoy}
\email{e.n.tsoy@gmail.com}
\affiliation{Physical-Technical Institute of the Uzbek Academy of Sciences,
Chingiz Aytmatov Str. 2-B, Tashkent, 100084, Uzbekistan}

\begin{abstract}
In this paper, we derive equations for the dynamics of ring dark solitons in an expanding cloud
of a two-dimensional Bose-Einstein condensate.  Assuming that the soliton's width
is much smaller than its radius,
we obtain the Hamilton equations for its evolution.  Then they are transformed into the Newton
equation, which is more convenient for applications.  The general theory is illustrated by the
solution of the Newton equation for the case of the axially symmetric condensate cloud, which
expands after switching off a harmonic trap.  The validity of our approximate analytical
approach is confirmed by comparison with the results of numerical simulations of
the Gross-Pitaevskii equation.
\end{abstract}



\maketitle

\section{Introduction}

Cylindrically symmetric nonlinear wave structures were first predicted and studied theoretically
in a two-dimensional generalization of the celebrated Korteweg-de Vries (KdV) equation
\cite{iord-59,ll-69,mv-74}. Thorough investigations of this cylindrical KdV equation showed
that it has ring-shaped soliton solutions provided the width of such solitons is much smaller
than the radius of the ring (see, e.g., \cite{hrs-23} and references therein). They were also
observed in experiments in Refs.~\cite{hr-74,NNK-78,step-81}. These solitons correspond to the
limit of small amplitude waves propagating in systems with weak dispersion. Since the curvature
of ring solitons is a small parameter, they can also be considered as solutions of the
Kadomtsev-Petviashvili (KP) equation \cite{kp-70} generalized to the case of dependence of the
wave variable on two polar coordinates in Ref.~\cite{johnson-80}. It is remarkable that all
these equations (KdV, cylindrical KdV, KP, and cylindrical KP) can be solved by the inverse
scattering transform method \cite{ll-69,dryma-74,dryma-76,dryma-83} and there are formal links
between solutions of different equations (see, e.g., Ref.~\cite{cms-07} and references therein).
These connections allowed the authors of Refs.~\cite{johnson-99,hzgs-24a} to demonstrate that a
localized pulse of a ring-like soliton is accompanied by a long small-amplitude tail, but the
evolution of the leading pulse is practically indistinguishable from an approximate adiabatic
dynamics of the KdV soliton with the account of a small curvature term.  Besides that, as is known,
soliton solutions of the KP equation in the form of a straight line are unstable with respect to
so-called ``snake'' instability \cite{kp-70,zakh-75}. Taking into account the dependence of the
wave variable on the azimuthal angle allowed one to study the stability of ring solitons with
respect to axial perturbations, and in case of instability a ring soliton evolves into a chain
of localized two-dimensional lumps \cite{hzgs-24a}.

If we go beyond an approximation of small-amplitude waves, when, for example, their dynamics
is described by a two-dimensional nonlinear Schr\"{o}dinger (NLS) equation in nonlinear optics
or a two-dimensional Gross-Pitaevskii (GP) equation in the physics of Bose-Einstein condensates
(BECs), then we arrive at similar phenomena.  Excitation of a dark soliton leads to the formation
of a counterflow shelf around the soliton even in one-dimensional geometry \cite{shevch-88}.
Such a shelf considerably changes the dynamics of narrow dark solitons under the action of external
forces when the GP equation becomes not completely integrable.  In particular, the effective mass
of a dark soliton quasiparticle is equal to 2 rather than 1 in standard non-dimensional units
\cite{ba-00,kp-04,pfk-05}. Dark ring solitons were observed in nonlinear optics (see, e.g.,
Refs.~\cite{sl-92,dnpgw-02}) and in BEC (see, e.g., Ref.~\cite{hoefer-06,tch-23}). If the width
of a ring-like soliton's profile is much smaller than the ring's radius, then the soliton's
curvature can be considered as a small perturbation \cite{ky-94,mss-11,kk-10}. The accuracy of
this approach to the description of dark soliton dynamics is confirmed by numerical simulations
(see, e.g., \cite{kk-10,skfs-11}). However, so far this method has been limited to solitons moving
through a not-flowing condensate confined in a stationary trap.  Another approach based on the
Whitham modulation theory (see Refs.~\cite{she-18,acehl-23}) describes the interaction of a soliton
with the background flow only for situations without the action of external forces and in
one-dimensional geometry.  Recently the Hamiltonian theory for the dynamics of dark solitons was
developed in Refs.~\cite{ik-22,ks-23,kamch-24a,ks-24,kamch-24b} and this method takes into account
automatically the effects of the counterflow.  Besides that, after the transformation of the Hamilton
equations to a more convenient Newton-like equation, one can easily take into account the external
forces provided they are weak enough and do not change the soliton's profile in the main approximation.
The resulting theory includes the effects of a large-scale background flow and agrees perfectly well
with the results of numerical simulations. The aim of this paper is to extend the theory to
situations with ring solitons evolving in the presence of an axially symmetric flow.

\section{Hamilton and Newton equations for dynamics of ring solitons}

We assume that dynamics of a ring soliton obeys the standard two-dimensional GP equation which can
be written in non-dimensional variables as
\begin{equation}\label{eq1a}
	i\psi_t+\frac12(\psi_{xx}+\psi_{yy})-|\psi|^2\psi=U(x,y)\psi,
\end{equation}
where $U(x,y)$ is the external potential and $\psi(x,y)$ is the condensate wave function. In our
axially symmetric geometry they only depend on the radius $r=(x^2+y^2)^{1/2}$ in polar
coordinates, that is we consider such stages of ring evolution, when one can neglect development
of the snake instability of dark solitons (see, e.g., \cite{kt-1988,ky-94,kk-10}). We suppose
that the width of soliton's profile is much smaller than the radius $R$ of the ring, so in the
main approximation we can neglect the curvature of the ring. Besides that, we confine ourselves
to situations with axially symmetric smooth potentials $U(r)$. Then a small arc of a ring can
be approximated by a straight segment and the soliton's profile is given by the known solution
\cite{tsuzuki-71} of the one-dimensional GP equation
\begin{equation}\label{eq1b}
	i\psi_t+\frac12\psi_{xx}-|\psi|^2\psi=U(x)\psi,
\end{equation}
where $x$ is the coordinate normal to the small arc.
It is convenient to transform this equation to hydrodynamic variables, namely, the condensate's
density $\rho$ and its flow velocity $u$, by means of the substitution
\begin{equation}\label{eq2}
	\psi(x,t)=\sqrt{\rho(x,t)}\exp\left({i}\int^x u(x',t)dx'\right),
\end{equation}
so that the GP equation (\ref{eq1b}) is cast to the system
\begin{equation}\label{eq3}
	\begin{split}
		&\rho_t+(\rho u)_x=0,\\
		&u_t+uu_x+\rho_x+\left[\frac{\rho_x^2}{8\rho^2}
		-\frac{ \rho_{xx}}{4\rho}\right]_x= -U_x.
	\end{split}
\end{equation}
Using the dark soliton solution of Eq.~(\ref{eq1b}) at $U(x) = 0$
\begin{equation}\label{eq4}
    \psi=\psi_s(x-Vt)=\left\{\sqrt{\rho_0-V^2}\tanh\left[\sqrt{\rho_0-V^2}\,(x-Vt)\right]+iV\right\}
    e^{-i\rho_0t},
\end{equation}
the distribution of the density and velocity of the soliton on the moving background
in the main approximation can be written as
\begin{equation}\label{eq5}
	\begin{split}
		\rho&=\rho_0-\frac{\rho_0-(V-u_0)^2}{\cosh^2\left[\sqrt{\rho_0-(V-u_0)^2}(x-Vt)\right]}, \\
		u&=V-\frac{\rho_0(V-u_0)}{\rho},
	\end{split}
\end{equation}
where $\rho_0$ and $u_0$ are the values of the hydrodynamic background variables at the
location of the soliton, and $V$ is the instant soliton's velocity (actually, this is the
speed of change of the ring radius, $V=dR/dt$).

As was shown by different methods in Refs.~\cite{shevch-88,ik-22,kamch-24a}, the canonical momentum
and the Hamiltonian per unit length of a straight soliton are given by the expressions
\begin{equation}\label{eq7}
\begin{split}
  p&=-2V\sqrt{\rho_0-(V-u)^2}+2\rho_0\arccos\frac{V-u_0}{\sqrt{\rho_0}},\\
  H&=\frac43\left[\rho_0-(V-u_0)^2\right]^{3/2}+u_0p.
  \end{split}
\end{equation}
To make a transition to similar variables for a ring soliton, we have to make the following
replacements (see, e.g., \cite{ky-94,mss-11,kk-10}). First, since we assume that a soliton
is narrow, then the background variables
change little across the soliton's width, so the variables $\rho_0,u_0$ must be replaced by
their local values at $r\approx R$, that is $\rho_0\to \rho(R,t), u_0\to u(R,t)$, where $R$
is the instant radius of the ring and $\rho(r,t),u(r,t)$ are solutions of the
dispersionless equations
\begin{equation}\label{eq8}
\begin{split}
\rho_t + (\rho u)_r+\frac{u\rho}{r} = 0, \quad
u_t + uu_r + \rho_r  = -U_r,
\end{split}
\end{equation}
obtained from the obvious two-dimensional generalization of Eqs.~(\ref{eq3}) in the limit of
small dispersion when the higher order derivatives can be neglected outside the soliton
location. Second, to take into account the soliton's curvature, we multiply Eqs.~(\ref{eq7})
by the soliton's instant length $2\pi R$ (the factor $2\pi$ can be omitted)
and arrive at the expressions
\begin{equation}\label{eq9}
\begin{split}
  p&=R\left\{-2\dot{R}\sqrt{\rho-(\dot{R}-u)^2}+2\rho\arccos\frac{\dot{R}-u}{\sqrt{\rho}}\right\},\\
  H&=\frac43R\left[\rho-(\dot{R}-u)^2\right]^{3/2}+up,
  \end{split}
\end{equation}
where we also made the replacement $V\to dR/dt\equiv\dot{R}$. These are the canonical momentum
and the Hamiltonian of a ring soliton. It is implied that the velocity $\dot{R}$ is excluded
from the Hamiltonian with help of the expression for $p$, so we have $H=H(p,R,t)$, where the
dependence on $R$ and $t$ appears via the background variables $\rho(R,t),u(R,t)$. Then the
ring soliton evolves according to the Hamilton equations
\begin{equation}\label{eq10}
   \frac{dR}{dt}=\left(\frac{\prt H}{\prt p}\right)_{R},\qquad
  \frac{dp}{dt}=-\left(\frac{\prt H}{\prt R}\right)_{p}.
\end{equation}

Elimination of velocity from the Hamiltonian is difficult to perform in practice, so it is more
convenient to transform the Hamilton equations (\ref{eq10}) to the Newton-like equation for
the radius $R=R(t)$ (see \cite{ik-22}). Differentiation of the momentum $p$ in Eq.~(\ref{eq9})
with respect to time $t$ gives
\begin{equation}\label{eq11}
  \begin{split}
  \frac{dp}{dt} &=\dot{R}\left\{-2\dot{R}\sqrt{\rho-(\dot{R}-u)^2}+2\rho\arccos\frac{\dot{R}-u}{\sqrt{\rho}}\right\}\\
  &+R\left\{-4(\ddot{R}-u_R\dot{R}-u_t)\sqrt{\rho-(\dot{R}-u)^2}+2(\rho_R\dot{R}+\rho_t)\arccos\frac{\dot{R}-u}{\sqrt{\rho}}\right\}.
  \end{split}
\end{equation}
For the derivative in the right-hand side of the second Eq.~(\ref{eq10}) we have
\begin{equation}\label{eq12}
  \begin{split}
  \left(\frac{\prt H}{\prt R}\right)_{p}=H^{(0)}+pu_R+R\left\{ \left.\frac{\prt H^{(0)}}{\prt\rho}\right|_{\dot{R},u}\rho_R
  + \left.\frac{\prt H^{(0)}}{\prt u}\right|_{\dot{R},\rho}u_R
  + \left.\frac{\prt H^{(0)}}{\prt \dot{R}}\right|_{\rho,u}\cdot\left.\frac{\prt \dot{R}}{\prt R}\right|_p\right\},
  \end{split}
\end{equation}
where
\begin{equation}\label{eq13}
  H^{(0)}=\frac43[\rho-(\dot{R}-u)^2]^{3/2}.
\end{equation}
The derivative $({\prt \dot{R}}/{\prt R})_p$ is to be calculated by differentiation of the
first expression (\ref{eq9}) with respect to $R$ with $p=\mathrm{const}$. After simple
calculations we obtain
\begin{equation}\label{eq14}
  \left.\frac{\prt \dot{R}}{\prt R}\right|_{p}=u_R-\frac{\dot{R}-u}{2R}
  +\frac{\rho+R\rho_R}{2R\sqrt{\rho-(\dot{R}-u)^2}}\arccos\frac{\dot{R}-u}{\sqrt{\rho}}.
\end{equation}
The other derivatives in Eq.~(\ref{eq12}) are easily calculated, so we get the expression for the
right-hand side of the second Eq.~(\ref{eq10}). Its comparison with Eq.~(\ref{eq11}) yields
\begin{equation}\label{eq15}
\begin{split}
  &2\ddot{R}-2u_t -\rho_R+(\dot{R}+u)u_R+(\dot{R}-u)\frac{u}{R}-\frac{2}{3R}\left[\rho-(\dot{R}-u)^2\right]\\
  &-\frac{\rho_t+(\rho u)_R+u\rho/R}{\sqrt{\rho-(\dot{R}-u)^2}}\arccos\frac{\dot{R}-u}{\sqrt{\rho}}=0.
  \end{split}
\end{equation}
After taking into account the continuity equation (\ref{eq8}), we obtain
\begin{equation}\label{eq16}
  2\ddot{R}-2u_t -\rho_R+(\dot{R}+u)u_R+(\dot{R}-u)\frac{u}{R}-\frac{2}{3R}\left[\rho-(\dot{R}-u)^2\right]=0.
\end{equation}
Here $\rho(R,t)$ and $u(R,t)$ are solutions of Eqs.~(\ref{eq8}) with $r$ replaced by the instant radius
of the soliton.

In the limit of large $R$ the last two curvature terms in Eq.~(\ref{eq16}) can be neglected
and we return to the equation
\begin{equation}\label{eq17}
  2\ddot{x}-2u_t-\rho_x+(\dot{x}+u)u_x=0
\end{equation}
for the one-dimensional dynamics of solitons, where $R(t)$ is replaced by the coordinate $x(t)$
(see Refs.~\cite{ba-00,ik-22}).

In case of condensate at rest with $u=0$, we reproduce the equation
\begin{equation}\label{eq18}
  2\ddot{R} -\rho_R-\frac{2}{3R}(\rho-\dot{R}^2)=0,
\end{equation}
which generalizes equations derived earlier for a particular case of a uniform condensate
\cite{mss-11} and a condensate confined in a harmonic trap \cite{kk-10}. Multiplication
of Eq.~(\ref{eq18}) by $\dot{R}$ and simple integration yield the integral of motion
\begin{equation}\label{eq19}
  \dot{R}^2-\rho=CR^{-2/3},
\end{equation}
where $C$ is an integration constant.

At last, we can eliminate $u_t$ from Eq.~(\ref{eq16}) with help of Eq.~(\ref{eq8}),
so we arrive at the equation
\begin{equation}\label{eq20}
  2\ddot{R}=-2U_R-\rho_R+(\dot{R}-u)\left(u_R-\frac{u}{R}\right)
  +\frac{2}{3R}\left[\rho-(\dot{R}-u)^2\right]
\end{equation}
convenient for practical applications of the theory. Let us illustrate it by a concrete example.

\section{Axially symmetric expansion of the background condensate}

Expanding clouds of BEC are often used in experiments (see, e.g., \cite{ps-03} and references
therein). At the initial moment of time the BEC is usually confined in a trap with a harmonic
potential, so the distribution of the density in 2D geometry and in the Thomas-Fermi
approximation has the form
\begin{equation}\label{eq21}
  \rho(r,0)=\left\{
  \begin{array}{ll}
  1-r^2/a^2,\qquad & r<a,\\
  0,\qquad  & r\geq a,
  \end{array}
  \right.
\end{equation}
where $a$ is the initial radius of the cloud. Naturally, the initial flow velocity is everywhere
equal to zero, $u(r,0)=0$. After switching off the trap potential $U$, the hydrodynamic variables
$\rho(r,t)$, $u(r,t)$ evolve according to Eqs.~(\ref{eq8}) with $U=0$. It is remarkable that
these equations have self-similar solutions (see, e.g., \cite{ovs-56,nemch-65,dyson-68,al-70}),
which in 2D axially-symmetric geometry take especially simple form
\begin{equation}\label{eq22}
  \rho(r,t)=\frac{a^2}{a^2+2t^2}\left(1-\frac{r^2}{a^2+2t^2}\right),\qquad
  u(r,t)=\frac{2rt}{a^2+2t^2},\qquad r<\sqrt{a^2+2t^2}.
\end{equation}
Actually, this form reflects the symmetry of Eq.~(\ref{eq2}) with respect to the so-called
`lens transformation' \cite{talanov-70,fibich-15}. In our non-dimensional variables, the soliton's
width has the order of magnitude of unity, so for applicability of the above theory to a soliton
evolving in such a cloud, we must satisfy the condition $a\gg1$. However, after transition to
the Newton equation (\ref{eq20}), the soliton is considered as an infinitely thin ring with
the radius $R(t)$, so we can normalize all spatial dimensions to the length $a$, so, to simplify the
notation, in solving the Newton equation we can assume $a=1$. Naturally, this dimensional parameter
should be restored in comparison of analytical formulas with numerical solutions of Eq.~(\ref{eq2})
(see below).

Thus, we assume that evolution of the BEC cloud is described by the formulas
\begin{equation}\label{eq23}
  \rho(r,t)=\left\{
  \begin{array}{ll}
  \frac{1}{1+2t^2}\left(1-\frac{r^2}{1+2t^2}\right),\qquad & r<1,\\
  0,\qquad  & r\geq 1,
  \end{array}
  \right.
\end{equation}
and
\begin{equation}\label{eq24}
  u(r,t)=\frac{2rt}{1+2t^2}.
\end{equation}
Substitution of these formulas into Eq.~(\ref{eq20}) with $U=0$ yields the equation
\begin{equation}\label{eq25}
  \frac{d^2R}{dt^2}=\frac23\frac{R}{(1+2t^2)^2}+\frac{1}{3R(1+2t^2)}
-\frac1{3R}\left(\frac{dR}{dt}-\frac{2Rt}{1+2t^2}\right)^2,
\end{equation}
which should be solved with the initial conditions
\begin{equation}\label{eq26}
  R(0)=R_0,\qquad \dot{R}=V_0,
\end{equation}
where $R_0$ is the initial radius of the ring and $V_0$ is the initial velocity of its
increase ($V_0>0$) or decrease ($V_0<0$). It is important that the initial velocity is
always smaller than the local sound velocity, $|V_0|<\sqrt{\rho(R_0,0)}=\sqrt{1-R_0^2}$,
that is the initial values must satisfy the condition
\begin{equation}\label{eq27}
  R_0^2+V_0^2\leq1.
\end{equation}
Thus, the dynamics of the ring soliton along the nonuniform and non-stationary
background~(\ref{eq21}) is reduced to a single second-order differential equation for $R(t)$.

To study solutions of Eq.~(\ref{eq25}), we will start with a particular case when
$\dot{R}=u(R,t)$, that is the soliton is always black during its motion and its relative
velocity with respect to the condensate is equal to zero. Then the last term in Eq.~(\ref{eq25})
vanishes and we easily get
\begin{equation}\label{eq28}
  R(t)=\frac12\sqrt{1+2t^2}.
\end{equation}
This particular solution corresponds to the initial conditions $R_0=1/2$ and $V_0=0$.

The solution (\ref{eq28}) suggests the substitution
\begin{equation}\label{eq29}
  R(t)=W(t)\sqrt{1+2t^2},
\end{equation}
so that Eq.~(\ref{eq25}) transforms to
\begin{equation}\label{eq30}
  \frac{d^2W}{dt^2}=-\frac{1}{3W}\left(\frac{dW}{dt}\right)^2-\frac{4t}{1+2t^2}\frac{dW}{dt}
  +\frac{1}{3(1+2t^2)}\left(\frac1W-4W\right).
\end{equation}
We introduce another independent variable $\tau=\tau(t)$ according to the definition
\begin{equation}\label{eq31}
  \tau(t)=\frac{1}{\sqrt{2}}\arctan(\sqrt{2}t),\qquad \frac{d\tau}{dt}=\frac{1}{1+2t^2},
\end{equation}
and obtain the equation
\begin{equation}\label{eq32}
  \frac{d^2W}{d\tau^2}=-\frac{1}{3W}\left(\frac{dW}{d\tau}\right)^2
  +\frac{1}{3}\left(\frac1W-4W\right).
\end{equation}
Its integration can be reduced to a quadrature by means of the replacement
\begin{equation}\label{eq33}
  W=y^{4/3},
\end{equation}
so Eq.~(\ref{eq32}) is cast to
\begin{equation}\label{eq34}
  \frac{9}{4}\frac{d^2y}{d\tau^2}=y^{-1/2}-4y,
\end{equation}
and this equation yields at once the integral
\begin{equation}\nonumber
  \frac{9}{16}\left(\frac{dy}{d\tau}\right)^2=y^{1/2}-y^2+C.
\end{equation}
The integration constant $C$ can be found from the initial conditions (\ref{eq26}),
so we get
\begin{equation}\label{eq35}
    \frac{9}{16}\left(\frac{dy}{d\tau}\right)^2=y^{1/2}-y^2-R_0^{2/3}(1-R_0^2-V_0^2).
\end{equation}
Returning to the variable $W$, we obtain
\begin{equation}\label{eq36}
  \left(\frac{dW}{d\tau}\right)^2=1-W^2-\left(\frac{R_0}{W}\right)^{2/3}(1-R_0^2-V_0^2).
\end{equation}
This first-order differential equation must be solved with the initial condition
\begin{equation}\label{eq37}
  W(0)=R_0,
\end{equation}
and the solution reads
\begin{equation}\label{eq38}
  \tau=\pm\int_{R_0}^W\frac{dW}{\sqrt{1-W^2-(R_0/W)^{2/3}(1-V_0^2-R_0^2)}}
\end{equation}
or, after replacement $W=z^{3/2}$,
\begin{equation}\label{eq39}
  \tau=\pm\frac32\int_{R_0^{2/3}}^{W^{2/3}}\frac{zdz}{\sqrt{z(1-z^3)-R_0^{2/3}(1-R_0^2-V_0^2)}},
\end{equation}
where sign is determined by the sign of the initial velocity $V_0$. Formulas (\ref{eq29}),
(\ref{eq31}), and (\ref{eq38}) or (\ref{eq39}) give the dependence of the soliton's radius
on time in implicit form.

It is clear from Eq.~(\ref{eq39}) that $z$ oscillates between two zeroes $z_1$ and $z_2$ of
the expression under the square root sign provided this expression is positive in this interval.
If $1-R_0^2-V_0^2=0$, then these zeroes are evident, $z_1=0$, $z_2=1$, and the expression is
positive between them. Due to inequality (\ref{eq27}), the graph of the general expression
is shifted down and the zeroes satisfy the inequalities
\begin{equation}\label{eq40}
  0<z_1<z_2<1
\end{equation}
for $0<1-R_0^2-V_0^2<1$. This means that in general case $W$ never vanishes as a function
of $\tau$, that is the ring soliton never shrinks to the center.

It is important, that according to Eq.~(\ref{eq31}) the interval $0\leq t<\infty$ corresponds
to the finite interval $0\leq\tau\leq \pi/(2\sqrt{2})$, that is the variable $W$ either
changes monotonously, or, after going through a few extrema, tends to some finite value
$W_{\infty}$ within the interval $z_1^{3/2}<W_{\infty}<z_2^{3/2}$. Consequently, at
asymptotically large time the radius of the ring increases linearly with time,
\begin{equation}\label{eq41}
  R(t)\approx \sqrt{2}\,W_{\infty}t,\qquad t\to\infty.
\end{equation}

In an important particular case with $V_0=0$ one of the zeroes is evident, $z_a=R_0^{3/2}$.
Then another zero can be found as a solution of a cubic equation and expressed in the form
\begin{equation}\label{eq42}
  z_b=-\frac{R_0^{2/3}}{3}+\frac{1}{2^{1/3}}\left[\left(\sqrt{\Delta}+1-\frac{20}{27}R_0^2\right)^{1/3}
  -\left(\sqrt{\Delta}-1+\frac{20}{27}R_0^2\right)^{1/3}\right],
\end{equation}
where
\begin{equation}\label{eq43}
  \Delta=\frac{1}{27}(27-40R_0^2+16R_0^4).
\end{equation}
It is easy to find that $z_1=z_a, z_2=z_b$ for $0<R_0<1/2$ and $z_1=z_b, z_2=z_a$ for $1/2<R_0<1$.
The plots of $W_a=z_a^{3/2}=R_0$ and $W_b=z_b^{3/2}$ as functions of $R_0$ intersect each other at
$R_0=1/2$, so the interval shrinks to a point and we return to the solution (\ref{eq28}). It is
remarkable that the function $W_b=z_b^{3/2}$ is approximated by the expression $W_b\approx1-R_0$
with accuracy better than 2.5\% what is enough for the most of applications.

Since we assume an infinitely thin ring solitons, the initial conditions can be, in principle,
chosen in such a way, that $1-R_0^2-V_0^2$ is as close to zero as we wish. In this formal limit,
Eq.~(\ref{eq38}) can be easily integrated explicitly and we obtain
\begin{equation}\label{eq44}
  W(\tau)=\sin(\arcsin R_0\pm\tau),
\end{equation}
that is
\begin{equation}\label{eq45}
  R(t)=\sin[\arcsin R_0\pm\arctan(\sqrt{2}t)/\sqrt{2}]\sqrt{1+2t^2},
\end{equation}
where the sign is determined by the sign of the initial velocity $V_0$. In case of positive $V_0$,
the argument of the $\sin$-function is always smaller than $\pi/2+\pi/2\sqrt{2}<\pi$, that is $W$
reaches at most one maximum and then tends to a constant value. In case of negative initial velocity,
the argument vanishes, if $R_0<\sin(\pi/2\sqrt{2})$, that is the radius of the ring can reach
however small values. If we denote $1-R_0^2-V_0^2=\eps\ll1$, then the smaller zero of the expression
under the square root in Eq.~(\ref{eq38}) is equal to $W_1\approx R_0\eps^{3/2}$. According to
Eq.~(\ref{eq44}), this value is reached at the moment  $\tau_1=\arcsin R_0$, that is the minimal
radius of the ring is equal to
\begin{equation}\label{eq46}
  R_{\text{min}}\approx\sqrt{1+2t_1^2}\,W_1\approx\frac{R_0\eps^{3/2}}{\cos(\sqrt{2}\arcsin R_0)}.
\end{equation}
Of course, in practice our approach loses its applicability when the radius
of the ring becomes of the order of magnitude of the soliton's profile width.

\begin{figure}[t]
\begin{center}
	\includegraphics[width = 8cm]{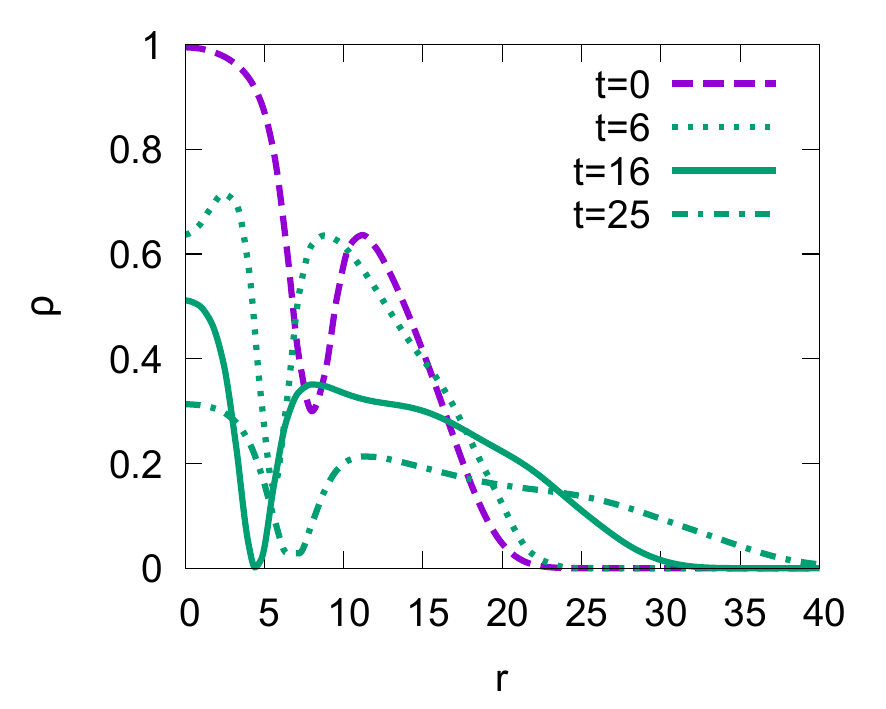}
\caption{Dependence of the condensate density on the radius $r$ for different moments of time.
The initial ring soliton radius is equal to $R=8$ and the
its initial `velocity' $V_0= -0.548$.
 }
\label{fig1}
\end{center}
\end{figure}

   In order to check our theory quantitatively, we compare numerical solutions of the 2D GP
equation (\ref{eq1a}) with solutions of the Newton equation (\ref{eq25}).
The initial radius of the condensate cloud
is equal to $a=20$, so it is much greater than the initial soliton width $\sim2$.
Correspondingly, all the parameters with a dimension of length in the above formulas,
e.g. in Eq.(\ref{eq25}),
must be measured in units of $a$, that is, we should make the replacements of the type
$R\to R/a$ and $t \to t/a$.

The initial distribution of the density is formed as the ground state of the
following potential
\begin{equation}\label{U_IC}
  U_{IC} = r^2 / a^2 + A \exp[-(r-R_0)^2 / (2 b^2)].
\end{equation}
Potential~(\ref{U_IC}) consists of a parabolic potential and a barrier, $A >0$. The parabolic
potential with $a \gg 1$ creates the density distribution close to that in Eq.~(\ref{eq21}).
The barrier with amplitude $A$ and width $b$ forms a dip with the minimum density
$V_0^2$, corresponding to the ring soliton. Parameters of the barrier, $A$ and $b$, are chosen
in such a way that the initial soliton has an appropriate width $\sim [(1 - R_0^2 /a^2) - V_0^2]^{-1/2}$,
see Eq.~(\ref{eq5}). The ground state of the potential~(\ref{U_IC}) is found numerically by the
Newton conjugate gradient method. Then, the phase distribution, according to the soliton
solution~(\ref{eq5}), is imposed to this ground state. The complex function, obtained by this
procedure, is taken as the initial condition $\psi(r, 0)$ for Eq.~(\ref{eq1a}). After formation of
the initial condition, potential~(\ref{U_IC}) is switched off.
The dashed line in Fig.~\ref{fig1} shows the dependence of the
density $\rho(r,0)$ on the radius $r$ with the ring dark soliton inserted at $R_0=8$.
The initial phase distribution corresponds to the negative initial velocity $V_0=-0.548$,
obtained at $A= 1$ and $b =0.80$ in Eq.~(\ref{U_IC}).
Numerical modeling of Eq.~(\ref{eq1a}) is performed by using the split-step Fourier method.
Distributions of the density, found by numerical simulations, at different
moments are shown in Fig.~\ref{fig1} by dotted, solid, and dash dotted lines.
As we see,  the condensate cloud expands according to the solution (\ref{eq22}).
At early stage of the cloud dynamics, linear waves are generated, see
the dotted line in Fig.~\ref{fig1} for $t= 6$, because the initial profile is not the exact
solution of Eq.~(\ref{eq1a}). These linear waves result in, for example, the density dip at
$r = 0$.  The ring radius decreases with time at the initial stage of evolution, and the
soliton becomes black with almost vanishing density at its center {\color{blue}near}
the turning point at $t \approx 16$, see the solid line in Fig.~\ref{fig1}.
After the turning point, the ring radius increases, see the dash-dotted line in Fig.~\ref{fig1}
for $t = 25$.

\begin{figure}[th]
\begin{center}
	\includegraphics[width = 8cm]{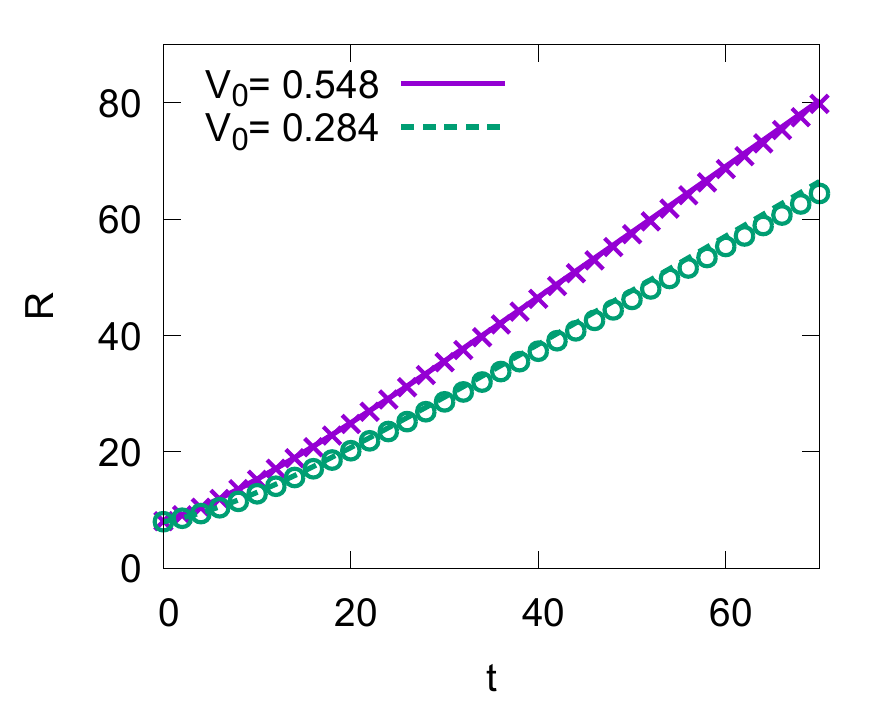}
\caption{Dependence of the soliton radius $R$ on time $t$ for two values of positive
initial velocity $V_0=0.284$ and $V_0=0.548$. Crosses and circles correspond to the numerical
solutions of the GP equation, solid and dashed lines to solutions of the Newton equation.
 }
\label{fig2}
\end{center}
\end{figure}

   We find the dependence the ring radius $R(t)$ on time $t$ from the numerical
solution of the 2D GP equation, and compare it with the solution of the Newton equation
with the same initial conditions.
In Fig.~\ref{fig2} we show the dependence  $R(t)$ for positive values
of the initial velocity.
The initial condition for $V_0 = 0.284$ is formed at $A= 5$ and $b =0.29$ in Eq.~(\ref{U_IC}).
The direct numerical solutions of the GP equation agree very well with
our approximate theory based on the Hamiltonian approach to the ring soliton dynamics.

\begin{figure}[th]
\begin{center}
	\includegraphics[width = 8cm]{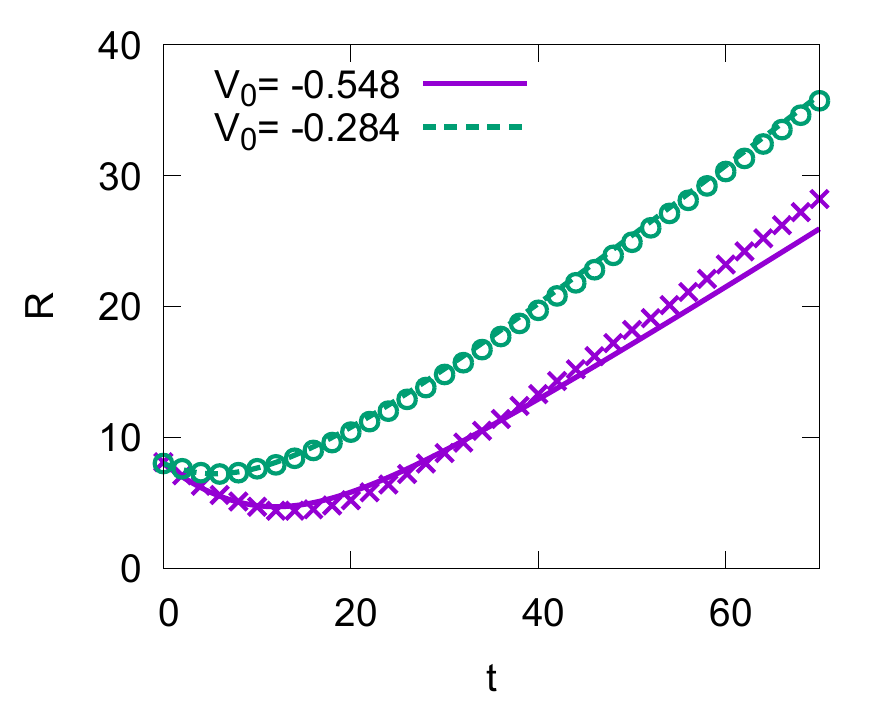}
\caption{Dependence of the soliton radius $R$ on time $t$ for two values of negative
initial velocity $V_0=-0.284$ and $V_0=-0.548$. Crosses and circles correspond to the numerical
solutions of the GP equation, solid and dashed lines to solutions of the Newton equation.
 }
\label{fig3}
\end{center}
\end{figure}

The dependence of the ring radius $R(t)$ on $t$ for negative initial velocities is shown in
Fig.~\ref{fig3}. Again, the agreement of direct numerical solutions of the GP equation with
solutions of the Newton equation is quite satisfactory.  The deviation of the analytical
solid line for $V_0=-0.548$ from the numerical crosses is apparently caused by the fact that
the minimal radius $R\sim5$ of the ring is not much greater than the soliton's width $\sim2$,
so the condition of applicability of the theory is not satisfied with very high accuracy.
Nevertheless, the accuracy better than 10\% even in this unfavorable situation for the theory
seems quite good.  Thus, the validity of the Hamiltonian approach to the dynamics of solitons
evolving along a non-uniform and time-dependent background is confirmed in this complicated
case of ring solitons.

\section{Conclusion}

Excitation of a dark soliton in BEC is accompanied inevitably by two effects: (a) atoms are
redistributed in such a way that the decrease of their density at the location of a soliton
must be compensated by the slight increase of their density in the background distribution so
that the total number of atoms is constant; (b) a moving dark soliton causes a local flow of
BEC in the vicinity of the soliton's location and, consequently, the jump of the condensate's
wave function phase across the soliton, so this jump must be compensated by the counterflow
in the background.  These two restrictions lead to considerable complications in calculations
of the canonical momentum and the Hamiltonian of a dark soliton by standard methods (see, e.g.,
Refs.~\cite{shevch-88,ba-00}). We suggested in Refs.~\cite{ks-23,kamch-24a,ks-24,kamch-24b} a
different approach based on the assumption that the soliton dynamics remain Hamiltonian even
when we separate approximately the evolution of a narrow soliton from the dispersionless
evolution of a smooth background.  This assumption seems very plausible for completely integrable
equations for which the Hamiltonian dynamics of solitons is well known in the case of a uniform
background \cite{zf-71,gardner-71}. Then, as was noticed in Refs.~\cite{ks-24c,kamch-25}, at
least for equations that belong to the AKNS scheme \cite{AKNS-74}, there exists an integral of
Hamilton equations governing the propagation of linear wave packets along a large-scale background.
With the use of Stokes remark \cite{stokes}, that linear waves satisfy the same equations as
small-amplitude tails of solitons, this integral can be converted into the integral for soliton
motion, and knowledge of such an integral allows one to find the canonical momentum and the
Hamiltonian for the soliton motion.

The formulated above approach is correct as long as the background is smooth enough and, hence,
in the main approximation, it can be considered constant within the soliton's location.
In a similar way, we can take into account the effects of external potentials provided the
gradients of the background variables resulting from their action are small enough.  In this paper,
we used the same idea and showed that, assuming that a small curvature of a soliton does not
affect its profile, we can obtain equations that govern the evolution of a ring soliton in BEC.
Actually, this leads to the addition of small curvature terms $\propto R^{-1}$ to the terms with
small gradients $U_R,\rho_R,u_R$ in the Newton equation (\ref{eq20}). It is clear that this method
is quite general, and it can be used in many other situations for taking into account additional
effects in the theory of soliton motion.

\begin{acknowledgments}

The research of AMK is funded by the research project FFUU-2021-0003 of the Institute of Spectroscopy
of the Russian Academy of Sciences (Sections~II) and by the RSF grant number~19-72-30028
(Section~III).

\end{acknowledgments}

\end{document}